\documentclass{article}[12]

\def\beq{\begin{equation}}
\def\eeq{\end{equation}}
\def\beqs{\begin{eqnarray}}
\def\eeqs{\end{eqnarray}}

\def\cg{{\mathbf   G}}

\def\cl{{\mathbf   L}}

\def\cs{{\mathbf   S}}

\def\ss{S^{1}}

\def\alg{\cl ie(G)}

\def\l2{L^{2}(\ss,\alg)}
\def\h1{H^{1}(\ss,G)}
\def\g0{\cg_{0}}

\def\eps{\epsilon}

\def\m#1{${#1}$}
\def\tr{{\rm tr}\;}
\def\half{{1\over 2}}
\def\pdr{\partial}
\def\un#1{\underline{#1}}

\begin{document}

\title{Three Dimensional Simplicial Yang-Mills Theory: An Approach to the Mass Gap}

\author{S.~G. ~Rajeev\\
Department of Physics and Astronomy \\
University of Rochester,Rochester, NY14627, USA\\ 
E-mail: rajeev@pas.rochester.edu}


\maketitle

\abstract{We show how to formulate Yang-Mills Theory in \m{2+1} dimensions
as a hamitonian system within  a simplicial regularization and construct its quantization, with special attention to the mass gap. An approximate  conformal invariance of the hamiltonian of this theory is useful to construct a
continuum limit.
}

\section{Introduction}
There has been substantial progress in recent years in understanding non-abelian Yang-Mills theories. The case of \m{1+1} dimensional Yang-Mills theory is particularly simple and has been solved in the lattice \cite{migdal}, hamiltonian \cite{rajeev} and path integral \cite{wittenfine} formalisms. 
We will study here the case of two space and one time dimension, where Karabali et. al. \cite{Karabalietal} have made remarkable progress; for example, they have computed the string tension
and given an argument that the `magnetic mass'  is non-zero. (The magnetic mass is a quantity that determines the decay of certain gluon correlation functions at spatial infinity). 

 It is our aim here  to compute another, more direct,  notion of a mass gap: the difference between the lowest two eigenvalues of the hamiltonian. We will do so rigorously within an approximation where the magnetic energy is ignored. It has been  argued heuristically \cite{feynmansinger} that the mass gap computed within this approximation  is in fact the exact answer for the complete theory. This reduced theory is invariant under conformal transformations of the spatial metric. Our regularization has a discrete version of this symmetry, which allows us to take the continuum limit.(A conformal symmetry also plays an important role  in the work of Karabali et al.) Our calculation gives answers that seem to agree with numerical simulations \cite{teper}.

 The analogous problem in three space and one time dimension is of course more interesting, but substantially more involved, because the ultra-violet divergences are worse; indeed it remains as one of the great challenges of mathematical physics \cite{claymath}.

\section{Hamiltonian Formulation of Yang-Mills Theory}
Let \m{G} be a semi-simple  Lie group and \m{X} a Riemann manifold of dimension \m{d}. We will
consider gauge fields as connections on the topologically trivial principal bundle
\m{X\times G}. These may be thought of as one-forms on \m{X} valued in
\m{\un{G}}.  If \m{G=R}, the abelian group of real numbers, the whole theory should reduce to electromagnetism.

A gauge transformation \m{g:X\to G} acts on such a
one-form 
\beq
A\mapsto gAg^{-1}+gdg^{-1}.
\eeq
 The curvature
\m{B(A)=dA+A\wedge A} transforms homogenously under this
transformation, and is the generalization of the magnetic field. The
Electric field is a \m{\un{G}}-valued \m{(d-1)}-form, canonically
conjugate to \m{A} (modulo some constraints we will list in a minute.)

The lagrangian of Yang--Mills theory may be written as  
\beq L=-\int \tr E*\left[{\pdr A\over \pdr t}-d_AA_0\right]+\half \int \tr[ \alpha E*E+{1\over \alpha}B(A)*B(A)] 
\eeq Here,
\m{A_0} is a \m{\un{G}}-valued scalar which generalizes the
electrostatic potential. Also, \m{\tr} denotes the trace in the
defining representation of \m{G} in \m{U(N)}, which induces a negative-definite
bilinear on \m{\un{G}}. The covariant exterior derivative is defined
as \m{ d_AA_0=dA_0+AA_0-A_0A.}

Variation with respect to \m{A_0} induces the constant on the pair
\m{(A,E)}, \m{ d_AE=0.}
These are first class constraints that encode the gauge invariance of the theory.
Following the usual philosophy of constrained dynamics, the true phase space of Yang-Mills theory thus consists of pairs \m{(A,E)} of Lie-algebra valued forms satisfying the above constraint, modulo the gauge equivalence relation
\m{
(A,E)\sim (gAg^{-1}gdg^{-1},gEg^{-1}).
}

The dynamics of the theory is given by the hamiltonian  
\beq
H=-\half \int \tr[\alpha E*E+{1\over \alpha}B(A)*B(A)] .
\eeq
The quantity \m{\alpha} is the gauge coupling constant; it measures the departure of the theory from being a free theory. For, we could re-scale  
\beq A\to \surd \alpha A, E\to {1\over \surd \alpha}E
\eeq
 without losing the fact that \m{A} and \m{E} are canonically conjugate. In the abelian theory, this constant would then drop out. But in the non-abelian theory it would multiply the terms in the hamiltonian that are not quadratic: \m{{1\over \alpha}B(A)*B(A)\to 
\left[dA+\surd \alpha [A,A]\right]^2}, thus becoming a measure of the strength of the non-linearity\footnote{
In  `natural' units in which \m{\hbar=c=1},   \m{\alpha} has the unit of mass for \m{d=2}.}.

\section{ Conformal and Symplectic  Invariance }

There are some special symmetries for this theory for low values of the dimension of space-time, which should be useful in obtaining some exact results. 

First of all, note that the constraint equations are invariant under all diffeomorphisms of space \m{X}. The metric tensor makes its appearance only in the formula for the hamiltonian. If \m{d=2}, the electric energy \m{-\int \tr E*E} is invariant under conformal transformations of the metric on \m{X}. To see this we just have to write it explicitly in tensor notation:
\m{
H_0=-\half\alpha\int \tr E*E=-\half\alpha\int \tr E_iE_j g^{ij}\surd gd^2x, 
}
the combination \m{\surd g g^{ij}} being invariant under \m{g_{ij}\to e^\phi g_{ij}}.
Thus this term in invariant under all conformal Killing vectors: an infinite dimensional Lie algebra for the case of two dimensional manifolds.

On the other hand, the magnetic energy (potential energy) 
\beq
V=-\half{1\over \alpha} \int\tr B(A)*B(A)
\eeq
 depends only on the area form induced by the metric:
the Hodge dual of a two form is a scalar obtained by contracting with the inverse of the area form, \m{*B=\half {1\over \surd g} \eps^{kl}B_{kl}}. Thus the magnetic energy is invariant all area-preserving transformations, another infinite dimensional Lie algebra. The total hamiltonian is invariant only under the intersection of these two Lie algebras, which is finite dimensional.

 T

\section{Reduced Yang--Mills Theory}

he hamiltonian divided by \m{\alpha} is dimensionless; the gauge coupling constant measures the relative strengths of the electric  and magnetic energies.

What if we ignore \m{V} and consider an approximation to Yang--Mills theory for which the hamiltonian is just \m{H_0} ( `Strong Coupling limit')? Since we preserve the non-abelian gauge invariance,we should expect many phenomena such as confinement to survive. More remarkably, it has been argued that \cite{feynmansinger} the mass gap of quantum Yang-Mills theory can be calculated exactly within this approximation, provided that it is non-zero: due to its scaling property, the potential energy will make a negligible contribution to this gap as the area of space tends to infinite.

To be a bit more concrete, suppose we start with an `Infra-Red regularized'  version of the quantum theory, in which \m{X} is compact but of large diameter \m{R} (more precisely, \m{\alpha R} is large).   If there are massless particles in the spectrum of \m{H_0} (`no gap'), we should expect that the 
difference between the lowest two eigenvalues of \m{H_0} will be of order  \m{R^{-2}}. For example, this is the case for the abelian theory with \m{G=R}. If there is a gap in the spectrum (so that the lightest particle in the truncated theory is massive), this difference will remain non-zero even if we scale the metric so that \m{\alpha R\to \infty}: the energy of the first excited state is the mass of the lightest particle. 

Under this scaling we expect the potential energy to behave like \m{(\alpha R)^{-2}}, so that it is negligible for large \m{R} {\em in the case where \m{H_0} has a gap}. No such conclusion can be made for the case of no gap. Also, the other energy differences  will behave like \m{R^{-2}} even when all particles are massive: for example, the energy of the second excited state minus the first excited state just corresponds to allowing the lightest particle to move with a small kinetic energy. Thus, the effect of \m{V} on them can not be ignored even for large \m{R}. Indeed \m{V} is the kinetic energy of these `glueballs'.

We will now show the reduced theory with hamiltonian \m{H_0} is an  exactly solvable quantum theory . Moreover, the mass-gap of this reduced theory  is  non-zero for  non-abelian gauge theories. We expect that the value of this quantity calculated within this approximation is exactly what it would be in Yang--Mills theory, although we do not provide a rigorous proof here.  In this approximation, the classical theory describes geodesics in the space of connections modulo gauge transformations. The metric induced on this space of gauge equivalence classes is curved and it is not at all a simple matter to understand its dynamics fully \cite{feynmansinger}. In the corresponding reduced quantum theory, the hamiltonian is the Laplacian on this infinite dimensional curved manifold. There have been attempts before to show that this manifold has positive sectional curvature which could imply the existence of a mass -gap. However, these approaches are plagued by various ultra-violet divergences, as is typical in quantum field theory.

In order to make any precise statement about a quantum field theory we need an ultra-violet regularization, preferably one that does not break its important symmetries. It looks at first impossible to regularize the reduced Yang--Mills theory without breaking scale invariance; for example a lattice regularization will introduce a constant that measures the distance between nearest neighbor points. We will instead  use a modification of the lattice regularization that does preserve a discrete version of conformal invariance; the cost will be that we must allow `lattices' that no longer have equal spacing between nearest neighbors. We will construct  a regularized version of 
 Yang--Mills theory  and  determine its spectrum   in the limit of ignoring the magnetic energy. Then we will show that the  gap is unchanged in the continuum limit. A discrete version of the conformal invariance (invariance under subdivision of simplices ) will make it possible to pass to the continuum limit.

\section{Simplicial Sets and Complexes}

Now let us build an analogue of Yang--Mills theory  on a discrete
approximation \m{\cs} to space \m{X}. A convenient language is
provided by the notion of simplicial set   \cite{petermay} in
algebraic topology. This is an abstraction of the more obvious notion of a triangulation 
of \m{X}. It will be convenient to formulate a discrete version of a gauge theory using only the abstract notion of a simplicial set.

A simplicial complex \m{\tilde K}  is a set of finite subsets of \m{X} (called simplices) of \m{X} such that every subset of an element of \m{\tilde K} is also an element of \m{\tilde K}.
We are to think of each set \m{\{x_0,x_1\cdots x_r\}} as an \m{r}-simplex (\m{r}-dimensional version of a triangular pyramid) with vertices at \m{x_0,x_1,\cdots x_r}. Thus, \m{0}-simplices are just points; \m{1}-simplices are to be thought of as lines (edges) connecting \m{\{x_0,x_1\}}, \m{2}-simplices are triangles (faces) with vertices at \m{\{x_0,x_1,x_2\}}, \m{3}-simplices are  tetrahedrons (volumes) with vertices at \m{\{x_0,x_1,x_2,  x_3   \}} and so on. A subset of a simplex in \m{\tilde K} is itself a simplex in it; thus the edges of a triangle or the faces of a tetrahedron in \m{\tilde K} are also in \m{\tilde K}. Because the vertices in a simplex are not ordered, we don't have any ordering of  the vertices. To deal with non-abelian gauge theories, we will need a slightly more refined notion that has such an ordering as well: a  {\em simplicial   set}, which we define next.

 Let \m{K_r} be the set of all sequences \m{(x_0,x_1,\cdots x_r)} such that the set \m{\{x_0,x_1\cdots x_r\}} is an \m{m}-simplex in \m{\tilde K} for some \m{m\leq r}. If some of the elements in the sequence are equal, the corresponding set will have fewer than \m{r} elements; then it will define a simplex of some lower dimension \m{m}.

There are maps \m{\pdr_i:K_r\to K_{r-1}} for \m{i=0,1,\cdots r} which pick out the \m{i}th face of \m{K_r} by omitting the \m{i}th vertex.
\beq
\pdr_i(x_0,\cdots x_r)=(x_0,x_1,\cdots, x_{i-1}, x_{i+1},\cdots x_r). 
\eeq

The face maps satisfy the relation
\beq
\pdr_i\pdr_j=\pdr_{j-1}\pdr_{i},\quad {\rm if}\; i<j \label{simpsetaxioms1}.
\eeq
 There are also certain maps \m{s_i:K_{r}\to K_{r+1}} 
 \beq
 s_i(x_0,\cdots x_r)=(x_0,x_1,\cdots x_i,x_i,\cdots x_r)
\eeq
 which are inverses of the face maps. (We don't need  them very much in what follows.) The  \m{K_r} together form a {\em simplicial set}, which means that the maps \m{\pdr_i,s_i} satisfy the axioms (easily verified although we don't use them)
\beqs
s_i s_j&=&s_{j+1}s_i,\quad {\rm if}\; i\leq j,\quad 
\pdr_is_j=s_{j-1}\pdr_i,\quad {\rm if}\; i<j,\cr
\pdr_js_j&=&{\rm identity}=\pdr_{j+1}s_j,\quad 
\pdr_is_j=s_j\pdr_{i-1},\quad {\rm if}\; i>j+1 \label{simpsetaxioms2}
\eeqs
in addition to (\ref{simpsetaxioms1}).

For example, an edge \m{e=(x_0,x_1)} has as its boundaries \m{\pdr_0e=x_1} and \m{\pdr_1e=x_0}, its ending  and starting  points. (Note that \m{\pdr_0} removes the starting vertex hence yielding  the ending vertex.) A face 
\m{f=(x_0,x_1,x_2)} is to be thought of as a triangle with vortices at \m{x_0,x_1,x_2}, and has boundaries which are the three edges of this triangle:
\beq
\pdr_0(x_0,x_1,x_2)=(x_1,x_2),\quad \pdr_1(x_0,x_1,x_2)=(x_0,x_2),\quad \pdr_2(x_0,x_1,x_2)=(x_0,x_1).
\eeq 
We can verify  \m{\pdr_0\pdr_2f=x_1=\pdr_1\pdr_0f} etc.

\section{Simplicial Gauge Theory}

We can now give the discrete version of a gauge theory on such  a simplicial set. 
A gauge transformation is a map \m{g:K_0\to G}. A connection is a map \m{U:K_1\to G}, satisfying the condition
\m{
U(x_0,x_1)=U(x_1,x_0)^{-1}.
}
The action of a gauge transformation on a  connection is
\m{
U(x_0,x_1)\mapsto g(x_0)U(x_0,x_1)g^{-1}(x_1).
}
In other words, 
\m{
U(e)\mapsto g(\pdr_1e)U(e)g(\pdr_0e)^{-1}.
}
The curvature of a connection 
is a map \m{\Phi:K_2\to G} given  by \m{\Phi(x_0,x_1,x_2)=U(x_0,x_1)U(x_1,x_2)U(x_2,x_0)} or equivalently, 
\m{
\Phi(f)=U(\pdr_2f)U(\pdr_0f)U(\pdr_1f)^{-1}.
}
 Under a gauge transformation,
\m{
\Phi(x_0,x_1,x_2)\mapsto g(x_0)\Phi(x_0,x_1,x_2) g^{-1}(x_0).
}
This can also be verified using just the axioms of  simplicial  sets:
\beqs
\Phi(f)&\mapsto & g(\pdr_1\pdr_2f)U(\pdr_2f)g(\pdr_0\pdr_2f)^{-1}g(\pdr_1\pdr_0f)U(\pdr_0f)g(\pdr_0\pdr_0f)^{-1}\cr
& & 
\left[g(\pdr_1\pdr_1f)U(\pdr_1f)g(\pdr_0\pdr_1f)^{-1}
\right]^{-1}\cr
&=&g(\pdr_1\pdr_2f)\Phi(f)g(\pdr_1\pdr_2f)^{-1}
\eeqs
using the relation (\ref{simpsetaxioms1}).
Thus  the curvature transforms in the adjoint representation at the starting point and hence \m{\tr\Phi(f) } is gauge invariant.
Moreover the curvature   satisfies the discrete analogue of the Bianchi identity:
\m{
\Phi(x_0,x_1,x_2)\Phi(x_0,x_2,x_3)\Phi(x_0,x_3,x_1)=U(x_0,x_1)\Phi(x_1,x_2,x_3)U(x_1,x_0),
}
which may also be written in the language of simplicial sets:
\beq
\Phi(\pdr_3v)\Phi(\pdr_1v)\Phi(\pdr_2v)=U(\pdr_2\pdr_3v)\Phi(\pdr_0v)U(\pdr_2\pdr_3v)^{-1},
\quad {\rm for}\; v\in K_3.
\eeq
The electric field is a map
\footnote{ It is more natural to let the Electric field take values in the dual of the Lie algebra; since we are using a non-degenerate invariant bilinear \m{\tr} throughout, we identify the Lie algebra with its dual.} 
  \m{E:K_1\to \un{G}} such that \m{E(x_0,x_1)=-E(x_1,x_0)}. We assign the gauge transformation 
\m{
E(x_0,x_1)\mapsto g(x_0)E(x_0,x_1)g(x_0)^{-1}
} [or alternately  \m{E(e)\mapsto {\rm ad}\; g(\pdr_1e)E(e)}], 
so that \m{\tr E^2} is invariant.

The lagrangian of simplicial gauge theory is then
\beqs
L&=&-\sum_{e\in K_1}\tr E(e)\left[{\pdr U(e)\over \pdr t}U(e)^{-1}+A_0(\pdr_1e)-U(e)A_0(\pdr_0e)U(e)^{-1}
\right]\cr
& & +\half\left[ a\sum_{e\in K_1} \tr E(e)^2+ b\sum_{f\in K_2} \tr \Phi(f) \right].
\eeqs
Here \m{a} and \m{b} are `coupling constants' which will have to be tuned to zero at some specific rate as we let 
the number of vertices edges and faces grow to infinity, in order to get the continuum limit. It is part of our 
 problem to determine these rates.

This Lagrangian is invariant under time dependent gauge transformations \m{g:K_0\times R\to G} acting  on the variables \m{U,E} as above  and on \m{A_0} as follows:
\m{
A_0(x)\mapsto g(x)A_0(x)g(x)^{-1}+g(x){\pdr\over \pdr t} g(x)^{-1}.
}

The constraint implied by varying \m{A_0} is a `conservation law' ( the discrete analogue of the divergence being zero):
\beq
\sum_{\pdr_1e=x}E(e)=\sum_{\pdr_0e=x} U(e)E(e)U(e)^{-1},\quad {\rm for}\; x\in K_0\label{constraint1}.
\eeq
It says that the sum of the electric fields over all edges starting at the vertex
 \m{x} is equal to the sum over all those ending at \m{x} ( except that we must parallel transport the latter using the connection to make
both sides transform the same way).

The equation of motion
\m{
aE(e)={\pdr U(e)\over \pdr t}U(e)^{-1}
}
obtained by varying \m{E(e)} shows that the electric field is the generator of left translations on the variables \m{U(e)}. 

Thus our system is analogous to a set of coupled rigid bodies, one on each edge of the simplicial set, with \m{E(e)} playing the role of angular momentum. The hamiltonian is  
\m{
H=\half\left[a\sum_{e\in K_1} \tr E(e)^2+ b\sum_{f\in K_2} \tr \Phi(f) 
\right].
}
The first term is the  kinetic energy and the second the  potential energy if we think of this as a set of coupled rigid bodies. This analogy allows us to pass to the quantum theory easily. 

\section{Quantization of Simplicial  Gauge Theory}
In the quantum theory,  
the  wave function is any function of the variables \m{U(e),e\in K_1 } satisfying the constraint of gauge invariance:
\beq
\psi\left(g(\pdr_1e)U(e)g(\pdr_0e )^{-1}\right)=\psi(U(e))\label{constraint2}.
\eeq
This is just the quantum analogue of the constraint (\ref{constraint1}) on the `angular momenta' \m{E(e)}. Also,
\m{
\hat H=\half a\sum_{e\in K_1} \Delta(e)+\half b\sum_{f\in K_2} \tr \Phi(f)
}
is the quantum hamiltonian, where \m{\Delta(e)} is the Laplace operator on the group \m{G}, acting on the variable \m{U(e)}.

By the Peter-Weyl theorem, we can expand the function of a compact Lie group in terms of the matrices of the irreducible representations of the group. Applied to our case, we find that a solution to the constraints above are given by the following process:

A.Decorate each edge \m{e} with an irreducible representation \m{\rho(e)} of \m{G}.

B.Attach an invariant map \m{\tilde\psi(x):\bigotimes_{\pdr_1e=x}\rho(e)\otimes \bigotimes_{\pdr_0e=x}\rho(e)^*\to C} to each vertex \m{x}.
The product of these \m{\tilde\psi(x)} defines a solution  of the \m{U(e)}. The most general solution of the constraint is a linear combination of these `invariant polynomials."

What we are doing is in fact a kind of non-abelian analogue of network theory. If the group \m{G} is the real line, the constraint (\ref{constraint1}) reduces to Kirchoff's law for conservation of current. Our condition (B) above is the non-abelian analogue of this conservation law: the tensor product of the incoming representations and the conjugate of the outgoing representations must contain the trivial representation.

The simplest solution is the `vacuum':  the function is a constant and all the representations \m{\rho(e)}  are trivial. The next simplest is
 to choose the representations to be trivial everywhere except on an edge loop that doesn't intersect itself anywhere, on which it is the fundamental representation. The trace of product of traces of the connection along this loop is an invariant (``Wilson loop'') and is a possible wave function.

Let us start to understand the spectral problem of the hamiltonian by ignoring the magnetic energy; we will recover this term in perturbation theory later:
\beq
H=H_0+H_1,\quad H_0=\half a\sum_{e\in K_1}\Delta(e), \quad H_1= \half b\sum_{f\in K_2} \tr \Phi(f)
\eeq
The ground state of the sum of \m{H_0} is simply the constant function. Simple representation theory shows that for  an edge loop \m{\gamma} that doesn't intersect itself anywhere,
\m{
W(\gamma)=\tr \prod_{e\in \gamma}U(e)
}
is an eigenfunction of the Laplace operator,
\m{
\sum_e\Delta(e)=C_2 l(\gamma)W(\gamma)
}
where \m{l(\gamma)} is the number of edges in the loop and \m{C_2} the quadratic Casimir of the defining representation.

Thus the first excited state of the Laplacian is the Wilson loop of the shortest length, formed by the edges around a face: nothing but \m{\tr \Phi(f)}:
\m{
H_0\tr\Phi(f)=3aC_2\tr \Phi(f).
}
There are as many such states as there are faces in our triangulation of \m{X}.

\section{ Subdivision of a face}

The discrete analogue of conformal invariance is the invariance under subdivision of the triangulation: we can subdivide a face without changing the rest of the triangulation, which corresponds to a local conformal transformation. Each triangulation induces a metric, the distance between two vertices being just the minimum number of edges in  a path connecting them. When we subdivide a particular face \m{f}, we change this metric only locally, without affecting  it for all the vertices not  in \m{f}. What is remarkable about the hamiltonian \m{H_0} is that its low lying eigenvalues are unchanged under such a subdivision, provided that the constant \m{a} is not changed: the ground state energy is still zero corresponding to the state where every link carries the trivial representation.
The first excited state will still be a loop around  any face, including possibly one of the new faces added by  the subdivision; its  energy will still be \m{3C_2a} where \m{C_2} is the quadratic Casimir of the fundamental representation.  In the dimensionless units we use, this  energy eigenvalue does not change. The degeneracy of the eigenvalue does change however, since now there are more faces. Indeed this is true of most of the the spectrum as a whole: the eigenvalues and eigenstates are unchanged under the subdivision, only the degeneracies change except when the size of the loop is comparable to the whole lattice. Because then, we might, by subdividing a face, create a loop of a  length longer than was possible before. But these are the states that will tend to have an energy of the order of the cut-off, and so we should ignore them. 

In this way we see that  the free energy
\m{
W(\beta)=-{1\over \#({\ {\rm faces}\ })}\log \tr e^{-\beta H_0}
}
tends to a finite limit as we subdivide faces repeatedly. Only  loops of length of order one will contribute substantially in this sum, as we take the limit of a large number of faces.  Their degeneracies are large ( of the order of the number of faces). Under each subdivision of a face, the change in these degeneracy is of order one, so in the limit  
it does not change the free energy. This is how we can construct the continuum limit of  reduced Yang--Mills  theory. As a corollary of this construction we see that the gap in the spectrum is \m{3C_2a} even in the continuum limit. The quantity \m{a} is just the gauge coupling constant of the continuum theory.

It may also be  possible to construct this continuum limit as a path integral on the space of connections modulo gauge transformations more  directly  along the lines of the construction of the volume measure of this space in  \cite{wittenfine}.

The eigenstates of \m{H_0} described above are nothing like the particles we expect in a quantum field theory: they describe loops (`glue rings') that are frozen in space. Of course we can take linear combinations of states with the same energy eigenvalue, to get other eigenstates, for example those that are momentum eigenstates. But the energy is independent of momentum; it is as if the particles are infinitely heavy. This is of course because we have ignored the magnetic energy. We can show that the magnetic energy of the Yang--Mills theory (which would be the potential energy in the language of gluons) serves as the kinetic energy term of these glue-rings. This should restore translation invariance and Lorentz invariance in the continuum limit, as well as justify the identification of the gap in the eigenvalues of \m{H_0} as the mass of the lightest particle in Yang-Mills theory. We will give a detailed argument in a later publication, noting here only that second order degenerate perturbation theory bears out this interpretation.



\section*{Acknowledgments}

I thank V. P. Nair  for discussions on the work cited above  and for reviving my interest in lower dimensional gauge theories. Also, I thank Rohana Wijewardhene for the opportunity to present this work at the Third Conference on Quantum Theory and Symmetries at the University of Cincinnati, Sep 10-14 2003.



\end{document}